\documentclass[apj]{emulateapj}  %for emulateapj style: *ONLY* put \begin{deluxetable}, \begin{figure*}  ,IF YOU WANT TO GO OVER TWO COLUMNS~~~

\slugcomment{Accepted by ApJ Letters on February 14, 2011}

\newcommand{\sm}{M$_{\odot}$}

\newcommand{\nsn}{47}  % total No of SN in this paper including published
\newcommand{\nsnnewspec}{35}  %No of SN in this paper with new host spectra presented here

\newcommand{\highestz}{0.18} %highest z of SN included in our sample for which we present data

\newcommand{\medianzIb}{0.015} %
\newcommand{\medianzIc}{0.017} %
\newcommand{\medianzIcbl}{0.043} %
\begin{document}

\title{Progenitor Diagnostics for Stripped Core-Collapse Supernovae: Measured Metallicities at Explosion Sites}
%\title{Measured Metallicities at the Sites of Stripped Core-Collapse
%Supernovae}

\author{M.~Modjaz\altaffilmark{1,2}, %
L.~Kewley\altaffilmark{3},
J.~S.~Bloom\altaffilmark{1,4}, 
A.~V.~Filippenko\altaffilmark{1}, 
D.~Perley\altaffilmark{1}, and
J.~M.~Silverman\altaffilmark{1,5}
}
\altaffiltext{1}{Department of Astronomy, University of California, Berkeley, CA 94720-3411; Miller Fellow.}
\altaffiltext{2}{Columbia Astrophysics Lab, Columbia University, NYC, NY 10024; Hubble Postdoctoral Fellow.}
\altaffiltext{3}{University of Hawaii, 2680 Woodlawn Drive, Honolulu, HI 96822.}
\altaffiltext{4}{Sloan Research Fellow.}
\altaffiltext{5}{Marc J. Staley Fellow.}

\begin{abstract}

  Metallicity is expected to influence not only the lives of massive
  stars but also the outcome of their deaths as supernovae (SNe) and
  gamma-ray bursts (GRBs). However, there are surprisingly few
  direct measurements of the local metallicities of different flavors
  of core-collapse SNe. Here we present the largest existing set of
  host-galaxy spectra with H~II region emission lines at the sites of
  \nsnnewspec\ stripped-envelope core-collapse SNe. We derive local
  oxygen abundances in a robust manner in order to constrain the
  SN~Ib/c progenitor population. We obtain spectra at the SN sites,
  include SNe from targeted and untargeted surveys, and perform
  the abundance determinations using three different oxygen-abundance
  calibrations. The sites of SNe~Ic (the demise of the most
  heavily stripped stars, having lost both H and He layers) are
  systematically more metal rich than those of SNe~Ib
  (arising from stars that retained their He layer) in all calibrations. A
  Kolmogorov-Smirnov test yields the very low probability of 1\% that SN~Ib and SN~Ic environment
  abundances, which are different on average by $\sim$0.2 dex (in the Pettini \& Pagel scale), are drawn from the same parent population. Broad-lined SNe~Ic (without GRBs) occur at metallicities between those of SNe~Ib
  and SNe~Ic. Lastly, we find that the host-galaxy central oxygen
  abundance 
  %, when inferred from the host-galaxy luminosity, 
  is not a
  good indicator of the local SN metallicity; hence, large-scale SN
  surveys need to obtain local abundance measurements in order to
  quantify the impact of metallicity on stellar death.

\end{abstract}

\keywords{galaxies: abundances--- supernovae: general}

\section{INTRODUCTION}\label{intro_sec}

Understanding the progenitors of the most energetic cosmic explosions,
particularly gamma-ray bursts (GRBs) and supernovae (SNe), is a grand
pursuit. Known already from their spectra is that stripped-envelope
core-collapse SNe (CCSNe; ``stripped SNe'' hereafter) have progressively
(from Types IIb to Ib to Ic) larger amounts of their outer hydrogen
and helium envelopes removed prior to explosion
\citep[e.g.,][]{clocchiatti96,filippenko97_review}. However, the
dominant mechanism of stripping is not well known, nor are basic
quantities such as the mass and metallicity of their stellar
progenitors. The exciting connection between long-duration GRBs and
broad-lined SNe~Ic (SNe~Ic-bl) and the existence of SNe~Ic-bl without
observed GRBs (see \citealt{woosley06_rev} for a review) raises the
question of what distinguishes a GRB progenitor from that of an
ordinary SN~Ic-bl without a GRB. Clear knowledge of the stellar
progenitors of various explosions is essential for understanding the
endpoints of stars over a broad mass range and for mapping the
chemical enrichment history of the universe \citep{nomoto06}.

Two progenitor channels have been proposed for stripped SNe: either
single massive Wolf-Rayet (WR) stars with main-sequence (MS) masses of
$\ga$ 30 \sm\ that have experienced mass loss during the MS and WR
stages (e.g., \citealt{fs85,woosley93}), or binaries from lower-mass
He stars that have been stripped of their outer envelopes through
interaction (\citealt{podsiadlowski04}, and references therein), or a
combination of both. Attempts to directly identify SN~Ib/c progenitors
in pre-explosion images have not yet been successful
(e.g., \citealt{gal-yam05,maund05}; \citealt{smartt09_rev}).

A more indirect but very powerful approach is to study the
environments of a large sample of CCSNe in order to discern systematic
trends that characterize their stellar populations. SNe~Ic tend to be
found in the brightest regions of their host galaxies \citep{kelly08}
and are more closely associated with H~II regions than SNe~II
(\citealt{anderson08}, and references therein).  SNe~Ib also tend to
be found in bright regions of their respective hosts, less closely
coupled than SNe~Ic but more so than SNe~II. This evidence suggests
the progenitors of SNe~Ib/c may thus be more massive than those of
SNe~II, which are $\sim 8$--16 \sm\ (see \citealt{smartt09_rev} for a
review). Other studies attempt to measure the metallicity by 
using the SN host-galaxy luminosity as a proxy
\citep{prantzos03,arcavi10}, or by using the metallicity of the galaxy
center measured from Sloan Digital Sky Survey (SDSS) spectra
\citep{prieto08} to extrapolate to that at the SN position
\citep{boissier09}.

Those prior metallicity studies do not directly probe the local
environment of each SN (which is different from the galaxy center due
to metallicity gradients), nor do they differentiate between the
different SN subtypes. Here we present a statistically significant
sample of stripped SNe (SNe~IIb and Ib, SNe~Ic, and SNe~Ic-bl) with
robust, uniform, and direct determinations of their local metallicity
in order to quantify the impact of metallicity on massive stellar
deaths, building on our previous work \citep{modjaz08_Z}. We note that
in the final stages of this research, \citet{anderson10} reported on a
similar topic; we briefly note the differences in the sample and
results below.

\section{The Supernova Sample and Associated Host Galaxies}\label{sample_sec}

In Table~\ref{sample_table}, we present the SN sample for which we
measured local metallicities. It consists of \nsnnewspec\ low-redshift
($z <$ \highestz ) stripped SNe, selected from the International
Astronomical Union Circulars
(IAUCs)\footnote{http://cfa-www.harvard.edu/iau/cbat.html .} according
to the following criteria: (1) well-determined SN subtype, (2)
discovered in targeted and untargeted surveys, and (3) H~II region
emission at the position of the SN (within our slit width of 1\arcsec,
see \S~\ref{specobs_sec}), as seen in our spectra, with which to
directly determine the metallicity at the SN site. We discuss the
potential impact of our selection effects in
\S~\ref{caveats_sec}. While our sample is heterogeneous and not
complete, we have good reasons to believe that it gives a fair
representation of the kinds of environments that give rise to observed
stripped CCSNe. Besides having SNe from traditional searches (e.g.,
the Lick Observatory SN Search; \citealt{filippenko01}) that target
luminous galaxies, we include SNe from untargeted surveys (e.g., SDSS,
Nearby SN Factory; see \citealt{modjaz08_Z} for more detailed
discussion) in order to mitigate any potential metallicity bias: since
SN host galaxies in targeted searches are preferentially more
luminous, they are usually also more metal rich \citep{tremonti04}.

Furthermore, we include host-galaxy spectra of the broad-lined SNe~Ic
without observed GRBs presented by \citet{modjaz08_Z}, which were
reduced and analyzed in the same fashion as the data presented here.
For completeness, we also include metallicity measurements of all
spectroscopically confirmed GRB-SNe (\citealt{modjaz08_Z}, and
references therein; \citealt{christensen08,chornock10,starling11}).

The total sample of SNe~Ib, Ic, and Ic-bl whose local metallicities we
are analyzing here amounts to \nsn\ SNe (without observed GRBs).

\section{Optical Spectroscopic Observations}\label{specobs_sec}

Optical long-slit ($1''$ wide) spectra of the locations of faded SNe
were obtained with the 10-m Keck~I telescope using the Low Resolution
Imaging Spectrometer \citep[LRIS;][]{oke95} plus atmospheric dispersion corrector (ADC) 
on a number of nights 2007--2010. We
generally employed a combination of the 300/5000 grism on the blue-side
CCD and the 400/8500 grating on the red-side CCD
%, to ensure sufficient
%resolution and broad wavelength range in order to cover the
%important emission lines. 
Here we also include our Keck LRIS/ADC
observations of stripped CCSNe in those cases where superimposed H~II
region emission lines were visible in the SN spectra \citep[e.g.,][]
{modjaz09,silverman09}.

All optical spectra were reduced and calibrated with standard
techniques in IRAF
%\footnote{IRAF is distributed by the National
%  Optical Astronomy Observatory, which is operated by the Association
%  of Universities for Research in Astronomy, Inc., under cooperative
 % agreement with the National Science Foundation (NSF).} 
 and our own
IDL routines for flux calibration \citep{matheson08}. For cases where
the SN was still present, we eliminated the SN contribution 
following the
successful method of \citet{modjaz08_Z}. After correcting all spectra
for their recession velocities we measured optical emission-line
fluxes by fitting Gaussians to the individual lines via the
$splot$ routine in IRAF. For the derivation of the statistical errors,
which typically amount to 5--10\% of the emission-line fluxes, we
follow \citet{perez03} and, in part, \citet{rupke10}.

\section{Metallicity Measurements}\label{metal_sec}

The nebular oxygen abundance is the canonical choice of metallicity
indicator for studies of the interstellar medium (ISM), since oxygen
is the most abundant metal, only weakly depleted, and exhibits very
strong nebular emission lines in the optical wavelength range
\citep[e.g.,][]{tremonti04}. Using our measured line fluxes of [O~II],
[O~III], [N~II], H$\alpha$, and H$\beta$, we correct for
reddening via the Balmer decrement and the standard Galactic
reddening law with $R_V = 3.1$ \citep{cardelli89}, and compute
the gas-phase oxygen abundance via strong-line diagnostics.

We employ three independent and well-known calibrations: (1) the
iterative method of \citet{kewley02}, as updated by \citet{kewley08}
(henceforth KD02-comb); (2) the calibration by \citet{mcgaugh91}
(henceforth M91); and (3) the diagnostic of \citet{pettini04} (both
PP04-O3N2 and N2), which is close to the direct electron temperature
($T_e$) scale. Moreover, we compute the uncertainties in the measured
metallicities by explicitly including the statistical uncertainties of
the line-flux measurements and those in the derived SN host-galaxy
reddening, and propagate them into the metallicity
determination. Since the PPO4-O3N2 scale utilizes ratios of lines that
are very close in wavelength, the effects of uncertain reddening and
scaling between the blue and red LRIS CCDs have negligible impact on
the abundance measurements (something we tested).  The independently
published metallicities of SNe 2006jc \citep{pastorello07}, 2007uy,
and 2008D \citep{thoene09} agree with our values within the
uncertainties.

%%%%%%%%%%%%% Cumulative Distribution plots l   %%%%%%%%%%%%%%%%%%%%%%%%
%\vspace{-0.5in}

\begin{figure}[!ht]
%\epsscale{0.5} %for manuscript
\centerline{\includegraphics[width=4in,angle=0]{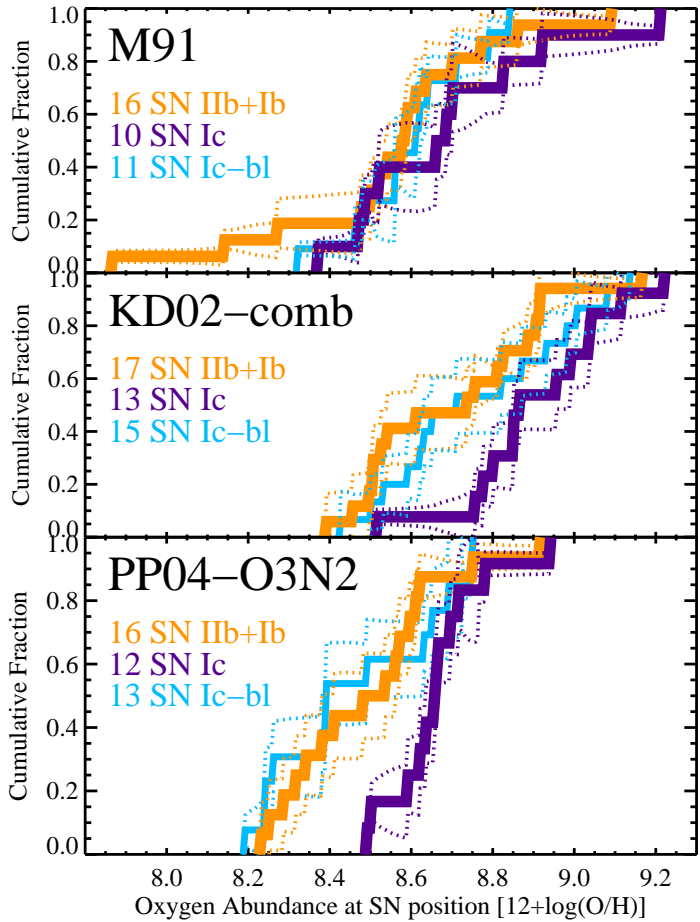}} %for laptop-emulateapj-
%\plotone{CCSN-LocalOxy_PP04_M91_cumdist_rev1.jpg}% for manuscript
%\vspace{-0.5in}
\caption{\singlespace Cumulative fraction (solid lines) of measured
  oxygen abundances at the SN position of different types of CCSNe
  with three different metallicity diagnostics and their confidence
  bands (dotted lines, see text). SNe~Ic (the demise of the most
  heavily stripped stars that lost much, if not all, of both their H
  and He layers) are systematically in more metal-rich environments
  than SNe~Ib (SNe arising from less stripped stars that retained
  their He layer). This is a robust conclusion since the
  trend is independent of the adopted metallicity diagnostic. 
  The PP04-O3N2 scale is least impacted by reddening and flux-calibration 
  uncertainties. Note that in this study, the SN~Ib subclass includes 
  SNe~IIb as well.
% Confidence bands were computed from 10,000 simulations 
%  via bootstrap with replacement.
  }
%%% THIS PLOT WAS MADE WITH /indirect/virgo4/mmodjaz/Ibc/Hostgals/spectra/plot_metall_ccsn,/SNGRB_in_cumdist
%%%%%%\rule{\textwidth}{0.5pt}
\label{ccsn_oxy_cumdist_fig}
\end{figure}
%%%%%%%%%%%%%%%%%%%%%%%%%%%%%%%%%%%%%%%%%%%

\section{Results}\label{results_sec}

Figure~\ref{ccsn_oxy_cumdist_fig} shows the cumulative distributions
of local metallicities in each of the three scales for different types
of stripped CCSNe: the ordinate indicates the fraction of the SN
population with metallicities less than the abscissa value. The SN
subtypes shown are SNe~Ib (including SNe~IIb), Ic, and Ic-bl (without
observed GRBs), and their respective numbers for which the requisite
emission lines for that diagnostic were available are given in the
legend. Furthermore, we show confidence bands around each cumulative
trend, which we computed via bootstrap with 10,000 realizations based
on our metallicity measurements and their associated
uncertainties. The metallicity measurements show well-known offsets
between different scales (e.g., \citealt{kewley08}); however, in 
{\it each} scale, SNe~Ic are more likely than SNe~Ib to be found in
metal-rich environments, while SNe~Ic-bl (without observed GRBs) have
environments that are similar in metallicity to those of both SNe~Ib
and SNe~Ic. 
A Kolmogorov-Smirnov (K-S) test reveals that the
probability that both the SN~Ib and SN~Ic local host-galaxy
metallicities have been drawn from the same parent population is low:
1\% (in PP04-O3N2), 7\% (in KD02-comb), and 15\% (in M91). The mean
local metallicities in the PP04-O3N2 scale are 12 + log(O/H) =
$8.49 \pm 0.19$ (for SNe~Ib), with a standard deviation of the mean (SDOM)
of 0.012; $8.66 \pm 0.12$  with SDOM of 0.010 (for SNe~Ic);
and $8.47 \pm 0.21$ with SDOM of 0.016  (for
SNe~Ic-bl). In all scales, SN~Ic sites have oxygen abundances
that are 0.14--0.20 dex higher than those of SN Ib.

Our results are somewhat at odds with those of \citet{anderson10}, who
find a slight and insignificant difference between environmental
metallicities of SNe~Ib and SNe~Ic in their sample from only targeted
SN searches. We have no SNe in common with their sample, and a
detailed comparison of the two different samples is required to
resolve the discrepancy, which is beyond the scope of this Letter.

\subsection{Tests for Possible Systematic Effects}\label{caveats_sec}

While there are selection effects that went into our heterogeneous
sample, none of them is expected to affect SNe~Ib systematically more
than SNe~Ic, and hence should not be able to cause our observed
trend. We checked that the SN survey mode does not explain the
observed trend; indeed, both SNe~Ib and SNe~Ic were drawn in almost
equal proportions from both targeted and untargeted surveys (Table 1),
and the relative difference in the metallicity of SN~Ib and SN~Ic
environments is visible even when only comparing SNe from the same
survey mode, albeit with more noise because of the smaller numbers of
objects.  Furthermore, we checked that both SNe~Ib and SNe~Ic span
comparable redshift ranges, with median redshifts of $\medianzIb$ (for
SNe~Ib) and $\medianzIc$ (for SNe~Ic). However, the broad-lined SNe~Ic
in our sample extend to larger redshifts, with a median of
$\medianzIcbl$.

While some surveys may have difficulty discovering SNe in the central
cores of bright galaxies, this detection difficulty does not appear to
affect one SN type more than another in our SN sample: from Table 1,
most of the SNe included here were found far from the central 1\arcsec, 
and those SNe that have offsets less than 1\arcsec\ comprise all
types (SN Ib, Ic, Ic-bl).  The only strong selection effect in our
sample is that we require H~II region emission lines to be present at
the SN position with which we can determine the ISM oxygen abundance,
meaning that the SN location has to have had a large amount of recent
(i.e., a few million years) star-formation activity. However, this
requirement affects SNe~Ib and SNe~Ic equally since our objects sample
the same redshift range.

\subsection{Supernova Progenitors with Metallicity-Driven Winds? }

A reasonable suggestion for why the environments of SNe~Ic are more
metal rich than those of SNe~Ib is that metallicity-driven winds
\citep{vink05,crowther06} in the progenitor stars prior to explosion
are responsible for removing most, if not all, of the He layer whose
spectroscopic nondetection distinguishes SNe~Ic from SNe~Ib. This
explanation may favor the single massive WR progenitor scenario as the
dominant mechanism for producing SNe~Ib/c \citep{fs85,woosley93}, at
least for those in large star-forming regions
(\S~\ref{caveats_sec}). While the binary scenario has been suggested
as the dominant channel for numerous reasons (see
\citealt{smartt09_rev} for a review; \citealt{smith11_snfrac}), we
cannot assess it in detail, since none of the theoretical studies
(e.g., \citealt{eldridge08}, and references therein) predict the
metallicity dependence of the subtype of stripped SN.  However, our
results are consistent with the suggestion of \citet{smith11_snfrac}
that SNe~Ic may come from stars with higher metallicities (and masses)
than SNe~Ib, even if they are in binaries.

The distribution of the SNe~Ic-bl is puzzling: SNe~Ic-bl seem to occur
at metallicities between those of SNe~Ib and SNe~Ic, except GRB-SNe
which are found at very low metallicities (but see
\citealt{soderberg10,levesque10_09bb}; also see
\citealt{levesque10_grb} for other GRB host metallicities). This may
indicate another key ingredient beyond metallicity for producing
ordinary SNe~Ic-bl, perhaps magnetic fields.

We find only a factor of 5 difference between the lowest metallicity
$Z$ for SNe~Ib and the highest metallicity for SNe~Ic. Since the
mass-loss rate $\dot{M}$ is proportional to $Z^{0.86}$ \citep{vink05},
this difference in $Z$ would imply a {\it maximum} factor of 4
difference in $\dot{M}$ between SN~Ib and SN~Ic progenitors. The
question remains whether this small difference in $\dot{M}$ is enough
to be responsible for removing all of the He layer, or other factors
are responsible that have a higher dependence on metallicity than
line-driven winds, or the metallicity trend simply correlates with
another property such as progenitor mass \citep{kelly08,anderson08}
that may determine the SN outcome. All observations and theoretical
work (see \citealt{bastian10} for a review) indicate that the initial
mass function is universal at the metallicities found herein.

Our results validate the independent hypothesis of \citet{arcavi10},
which was based on indirect data: as an explanation for the fact that
none of the 15 CCSNe found by the Palomar Transient Factory (PTF) in
dwarf galaxies ($M_R < -18$ mag) was a SN~Ic, \citet{arcavi10} suggest
that SNe~Ic do not occur at low metallicity since low-luminosity
galaxies usually have low metallicity, in contrast to SNe~Ic-bl of
which two were found by PTF in dwarf hosts. Here we have supporting
direct evidence; we show that SN~Ic host environments have
systematically higher metallicities than those of SNe~Ib, while those
of SNe~Ic-bl encompass both low and high abundances. Nevertheless, it
is important to measure metallicities directly and not rely on the
host-galaxy luminosity ($L$) as a proxy, as we show next.

%%%%%%%%%%%%% Local Z vs M_B   %%%%%%%%%%%%%%%%%%%%%%%%
\begin{figure}[!ht]
%\epsscale{0.5} %for manuscript
%\centerline{\includegraphics[width=3.75in,angle=90]{CCSN-MBvslocalZ.ps}} %for emulateapj
%\centerline{\includegraphics[width=6.75in,angle=90]{CCSN-MBvslocalZ.ps}} %for manuscript
\centerline{\includegraphics[width=4.in,angle=0]{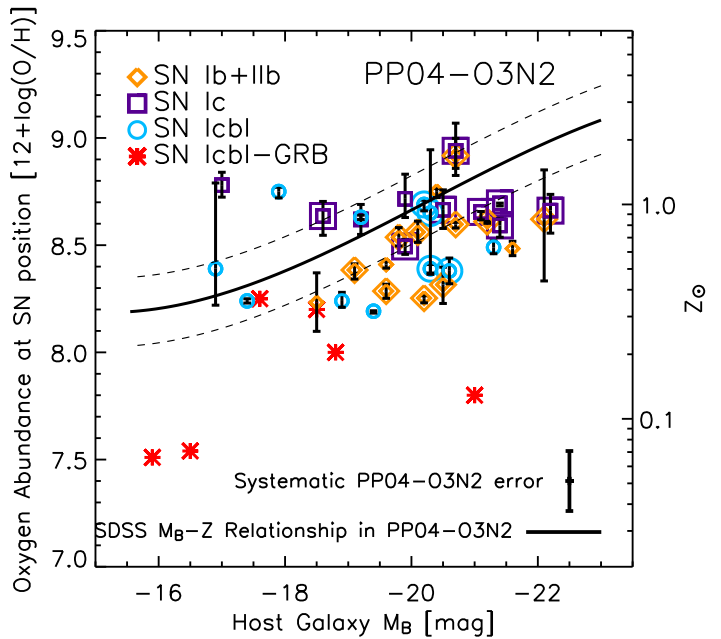}} %for laptop-emulateapj-

\caption{Measured oxygen abundance at the position of different SN
  types vs. the SN host-galaxy luminosity, and for comparison the
  oxygen abundance as inferred from the SDSS $L$--$Z$ relationship
  \citep{tremonti04}, converted to PP04-O3N2 (solid line) including
  $1\sigma$ uncertainties (dashed lines). The nuclear metallicity as
  derived from the SN host luminosity using the SDSS relationship is
  not a good proxy for the local oxygen abundance of SN
  environs. Here, we adopt a solar oxygen abundance of 12 + log(O/H) =
  8.69 \citep{asplund09_rev}, but note that the exact value of the
  solar abundance has no impact on our conclusions.
%  The environments of GRB-SNe have systematically lower chemical
%  abundances than all other samples of CCSN environments. 
  SNe found in targeted SN surveys are designated by extra circles,
  squares, and diamonds. The representative systematic error of 0.14
  dex for the PP04-O3N2 scale is shown in the bottom-right corner.
%  Our metallicity sample is larger than that
%  shown, since the $M_B$ value has not yet been measured for a number
%  of our SN host galaxies from untargeted surveys. 
}

%%%%%%\rule{\textwidth}{0.5pt}
\label{ccsn_MBvslocalZ_fig}
\end{figure}
%%%%%%%%%%%%%%%%%%%%%%%%%%%%%%%%%%%%%%%%%%%

\subsection{The Need for Local Metallicity Measurements}\label{local_sec}

Since we possess direct local metallicity measurements, here we test whether
SN host-galaxy luminosity and measured nuclear metallicity is a good proxy for the local SN
metallicity, as assumed in some studies.
%by simulating their technique
%and comparing the outcome to our direct measurements. 
To that end, we
drew the SN host-galaxy luminosities from the Lyon-Meudon
Extragalactic Database (HyperLEDA)\footnote{http://leda.univ-lyon1.fr/
  .} as their sample has been homogeneously compiled, and adopt their
reported absolute $B$-band magnitudes, $M_B$, in Table 1. Note that
the lowest-luminosity SN host galaxies are from untargeted surveys. Furthermore, 14 galaxies in our sample have nuclear SDSS spectra and their PP04 metallicities were taken from SDSS\footnote{http://www.mpa-garching.mpg.de/SDSS/DR7}. 

In Figure~\ref{ccsn_MBvslocalZ_fig} we plot the measured oxygen
abundance (on the PP04-O3N2 scale) at the position of stripped CCSN
vs.  the SN host-galaxy luminosity, and for comparison the oxygen
abundance as would be inferred from the SDSS $L$--$Z$ relationship
\citep{tremonti04} including $1\sigma$ uncertainties. For consistency,
we have converted the $L$--$Z$ relationship from \citet{tremonti04} to
the scale of PP04-O3N2 using the empirical calibrations of
\citet{kewley08}. Figure~\ref{ccsn_MBvslocalZ_fig} demonstrates that
nuclear metallicity when derived from the SN host luminosity is not a
good proxy for the local oxygen abundance of the environments of SNe:
the local metallicities are often lower than the inferred central ones
(because of metallicity gradients; \citealt{vanzee98}), but also
occasionally larger (e.g., \citealt{young10}), and in most cases more
deviant than the $1\sigma$ metallicity uncertainties (i.e., $0.16$
dex) would indicate from SDSS. The differences between the predicted
central and the measured local metallicity values range from $-0.4$
dex to $+0.5$ dex (up to $3\sigma$ away from the SDSS $L$--$Z$ value),
with a median of $0.2$ dex. Moreover, the nuclear PP04-O3N2 metallicities are deviant from the local values by 
an average of 0.13 dex, with extremes as large as 0.24 dex, and always more deviant than our measured uncertainties. Thus, we conclude that nuclear metallicities are not a good measure of local metallicities.

\section{Conclusions}\label{conclusion_sec}

We present a statistically significant sample of \nsnnewspec\ new
host-galaxy spectra of stripped CCSN (SNe~IIb and Ib, SNe~Ic, and
SNe~Ic-bl), and in combination with published host-galaxy spectra, we
perform robust, uniform, and direct determination of their local
metallicity. The aim is to search for metallicity trends that may let
us differentiate between various debated SN~Ib/c progenitor scenarios.
We find that the environments of SNe~Ic are systematically more metal
rich in three scales than those of SNe~Ib (on average by
 0.14--0.20~dex in all scales), with a K-S test yielding very small
probabilities that they are drawn from the same parent population. We
also show that SNe~Ic-bl (without GRBs) are intermediate to those other
classes.

For the future, we recommend this kind of detailed local metallicity
study for all subtypes of CCSNe from galaxy-impartial, large, and deep
photometric surveys (e.g., PTF, Pan-STARRS, Skymapper, LSST) in order
to comprehensively understand the impact of metallicity on massive
stellar death, as well as to compute cosmologically important
parameters, such as SN rates, as a function of metallicity.

{\it Note added in proof}: After our manuscript was accepted,
\citet{leloudas11} appeared on the arXiv-server on a similar
topic. While we corrected one SN identification based on their work, we are not
able to make more comparisons at this time.

%%%%%%%%%%%%%%%%%%%%%%%%%%%%%%%%%%

\acknowledgements

We thank the referee, J. Brinchmann, for helpful comments that improved the manuscript, and R. C. Thomas and J. Marriner for refining via SNID
(\citealt{blondin07}) the subtypes of some of the CCSNe included in this study.
We thank R. J. Foley, R. Chornock, T. N. Steele, and
D. Poznanski for helping to obtain some spectra, and D. Rupke and
N. Smith for useful discussions. M.M. acknowledges support from the
Miller Institute at UC Berkeley and from Hubble Fellowship grant
HST-HF-51277.01-A, awarded by STScI, which is operated by AURA under
NASA contract NAS5-26555.  L.K. acknowledges support from NSF Early
CAREER award AST--0748559.  A.V.F.'s group is grateful for financial
assistance from the TABASGO Foundation and NSF grant
AST--0908886. J.S.B. and D.A.P. were partially supported by a DOE
SciDAC grant and HST-GO-11551.
%(``When Good Stars Go Bang"; S. Woosley, PI).
%see {\t thttp://www.scidac.gov/physics/grb.html}).
The data presented herein were obtained at the W.~M. Keck Observatory,
which is operated as a scientific partnership among the California
Institute of Technology, the University of California, and NASA; the
observatory was made possible by the generous financial support of the
W. M. Keck Foundation.

%We wish to extend special thanks
%to those of Hawaiian ancestry on whose sacred mountain we are
%privileged to be guests.

%Furthermore, this research has made use of NASA's Astrophysics Data
%System Bibliographic Services (ADS), the HyperLEDA database, and the
%NASA/IPAC Extragalactic Database (NED) which is operated by the Jet
%Propulsion Laboratory, California Institute of Technology, under
%contract with NASA. 

%%%%%%%%%%%%%%%%%%%%%%%%%%%%%%%%%%%%%%%%%%%
%\clearpage
%%%%%%%%%%%%%%%% BIBLIOGRAPHY  %%%%%%%%%%%%%%%%%%%%%%%% 
%\bibliographystyle{apj}
%\bibliography{/indirect/virgo4/mmodjaz/Ibc/refs}

\begin{thebibliography}{50}
\expandafter\ifx\csname natexlab\endcsname\relax\def\natexlab#1{#1}\fi


\bibitem[{{Anderson} {et~al.}(2010){Anderson}, {Covarrubias}, {James}, {Hamuy},
  \& {Habergham}}]{anderson10}
{Anderson}, J.~P., {Covarrubias}, R.~A., {James}, P.~A., {Hamuy}, M., \&
  {Habergham}, S.~M. 2010, \mnras, 407, 2660 

\bibitem[{{Anderson} \& {James}(2008)}]{anderson08}
{Anderson}, J.~P., \& {James}, P.~A. 2008, \mnras, 390, 1527

\bibitem[{{Arcavi} {et~al.}(2010){Arcavi}, {Gal-Yam}, {Kasliwal}, {Quimby},
  {Ofek}, {Kulkarni}, {Nugent}, {Cenko}, {Bloom}, {Sullivan}, {Howell},
  {Poznanski}, {Filippenko}, {Law}, {Hook}, {Jonsson}, {Blake}, {Cooke},
  {Dekany}, {Rahmer}, {Hale}, {Smith}, {Zolkower}, {Velur}, {Walters},
  {Henning}, {Bui}, {McKenna}, \& {Jacobsen}}]{arcavi10}
{Arcavi}, I., et~al.\ 2010, \apj, 721, 777 


\bibitem[{{Asplund} {et~al.}(2009){Asplund}, {Grevesse}, {Sauval}, \&
  {Scott}}]{asplund09_rev}
{Asplund}, M., {Grevesse}, N., {Sauval}, A.~J., \& {Scott}, P. 2009, \araa, 47,
  481


\bibitem[{{Bastian} {et~al.}(2010){Bastian}, {Covey}, \& {Meyer}}]{bastian10}
{Bastian}, N., {Covey}, K.~R., \& {Meyer}, M.~R.\ 2010, \araa, 48, 339 


\bibitem[{{Blondin} \& {Tonry}(2007)}]{blondin07}
{Blondin}, S., \& {Tonry}, J.~L. 2007, \apj, 666,1024


\bibitem[{{Boissier} \& {Prantzos}(2009)}]{boissier09}
{Boissier}, S., \& {Prantzos}, N. 2009, \aap, 503, 137

%\bibitem[{{Bresolin} {et~al.}(2009){Bresolin}, {Gieren}, {Kudritzki},
 % {Pietrzy{\'n}ski}, {Urbaneja}, \& {Carraro}}]{bresolin09}
%{Bresolin}, F., {Gieren}, W., {Kudritzki}, R., {Pietrzy{\'n}ski}, G.,
  %{Urbaneja}, M.~A., \& {Carraro}, G. 2009, \apj, 700, 309

\bibitem[Cardelli et al.(1989)]{cardelli89} Cardelli, J.~A., 
Clayton, G.~C., \& Mathis, J.~S.\ 1989, \apj, 345, 245 

\bibitem[{{Chornock} {et~al.}(2010){Chornock}, {Berger}, {Levesque},
  {Soderberg}, {Foley}, {Fox}, {Frebel}, {Simon}, {Bochanski}, {Challis},
  {Kirshner}, {Podsiadlowski}, {Roth}, {Rutledge}, {Schmidt}, {Sheppard}, \&
  {Simcoe}}]{chornock10}
{Chornock}, R., et~al.\ 2010, ApJ, submitted, (arXiv:1004.2262)

\bibitem[Christensen et al.(2008)]{christensen08} Christensen, L., Vreeswijk, P.~M., Sollerman, J., Th{\"o}ne, C.~C., Le Floc'h, E., \& Wiersema, K.\ 2008, \aap, 490, 45 



\bibitem[{{Clocchiatti} {et~al.}(1996){Clocchiatti}, {Wheeler}, {Brotherton},
  {Cochran}, {Wills}, {Barker}, \& {Turatto}}]{clocchiatti96}
{Clocchiatti}, A., {Wheeler}, J.~C., {Brotherton}, M.~S., {Cochran}, A.~L.,
  {Wills}, D., {Barker}, E.~S., \& {Turatto}, M. 1996, \apj, 462, 462

%\bibitem[{{Crockett} {et~al.}(2008){Crockett}, {Eldridge}, {Smartt},
 % {Pastorello}, {Gal-Yam}, {Fox}, {Leonard}, {Kasliwal}, {Mattila}, {Maund},
 % {Stephens}, \& {Danziger}}]{crockett08_08ax}
%{Crockett}, R.~M., et~al.\ 2008, \mnras,  391, L5

\bibitem[{{Crowther} \& {Hadfield}(2006)}]{crowther06}
{Crowther}, P.~A., \& {Hadfield}, L.~J. 2006, \aap, 449, 711

\bibitem[{{Eldridge} {et~al.}(2008){Eldridge}, {Izzard}, \&
  {Tout}}]{eldridge08}
{Eldridge}, J.~J., {Izzard}, R.~G., \& {Tout}, C.~A. 2008, \mnras, 384, 1109

\bibitem[{{Filippenko}(1997)}]{filippenko97_review}
{Filippenko}, A.~V. 1997, \araa, 35, 309

\bibitem[{Filippenko} {et~al.}(2001)]{filippenko01} {Filippenko}, A. V.,
  {Li}, W. D., {Treffers}, R. R., \& {Modjaz}, M. 2001, in 
   Small-Telescope Astronomy on Global Scales, ed.
   W. P. Chen, C. Lemme, \& B. Paczy\'{n}ski (San Francisco: ASP), 121

\bibitem[{{Filippenko} \& {Sargent}(1985)}]{fs85}
{Filippenko}, A.~V., \& {Sargent}, W.~L.~W. 1985, \nat, 316, 407

%\bibitem[{{Fryer} {et~al.}(2007){Fryer}, {Mazzali}, {Prochaska}, {Cappellaro},
 % {Panaitescu}, {Berger}, {van Putten}, {van den Heuvel}, {Young},
  %{Hungerford}, {Rockefeller}, {Yoon}, {Podsiadlowski}, {Nomoto}, {Chevalier},
 % {Schmidt}, \& {Kulkarni}}]{fryer07}
%{Fryer}, C.~L., et~al.\ 2007, \pasp, 119, 1211

\bibitem[{{Gal-Yam} {et~al.}(2005){Gal-Yam}, {Fox}, {Kulkarni}, {Matthews},
  {Leonard}, {Sand}, {Moon}, {Cenko}, \& {Soderberg}}]{gal-yam05}
{Gal-Yam}, A., et~al.\ 2005, \apj, 630, L29

%\bibitem[{{Gaskell} {et~al.}(1986){Gaskell}, {Cappellaro}, {Dinerstein},
%  {Garnett}, {Harkness}, \& {Wheeler}}]{gaskell86}
%{Gaskell}, C.~M., {Cappellaro}, E., {Dinerstein}, H.~L., {Garnett}, D.~R.,
%  {Harkness}, R.~P., \& {Wheeler}, J.~C. 1986, \apj, 306, L77

\bibitem[{{Kelly} {et~al.}(2008){Kelly}, {Kirshner}, \& {Pahre}}]{kelly08}
{Kelly}, P.~L., {Kirshner}, R.~P., \& {Pahre}, M. 2008, \apj, 687, 1201

\bibitem[{{Kewley} \& {Dopita}(2002)}]{kewley02}
{Kewley}, L.~J., \& {Dopita}, M.~A. 2002, \apjs, 142, 35

\bibitem[{{Kewley} \& {Ellison}(2008)}]{kewley08}
{Kewley}, L.~J., \& {Ellison}, S.~L. 2008, \apj, 681, 1183

\bibitem[Leloudas et al.(2011)]{leloudas11} Leloudas, G., et al.\ 
2011, A \& A, submitted (arXiv:1102.2249) 

%\bibitem[{{Kobulnicky} \& {Kewley}(2004)}]{kobulnicky04}
%{Kobulnicky}, H.~A., \& {Kewley}, L.~J. 2004, \apj, 617, 240

\bibitem[Levesque et al.(2010b)]{levesque10_grb} Levesque, E.~M., 
Kewley, L.~J., Berger, E., \& Jabran Zahid, H.\ 2010b,  \aj, 140, 1557

\bibitem[{{Levesque} {et~al.}(2010a){Levesque}, {Soderberg}, {Foley}, {Berger},
  {Kewley}, {Chakraborti}, {Ray}, {Torres}, {Challis}, {Kirshner}, {Barthelmy},
  {Bietenholz}, {Chandra}, {Chaplin}, {Chevalier}, {Chugai}, {Connaughton},
  {Copete}, {Fox}, {Fransson}, {Grindlay}, {Hamuy}, {Milne}, {Pignata},
  {Stritzinger}, \& {Wieringa}}]{levesque10_09bb}
{Levesque}, E.~M., et~al.\ 2010, \apj,  709, L26

%\bibitem[{{Li} {et~al.}(2007){Li}, {Wang}, {Van Dyk}, {Cuillandre}, {Foley}, \&
  %{Filippenko}}]{li07_SNIIprog}
%{Li}, W., {Wang}, X., {Van Dyk}, S.~D., {Cuillandre}, J., {Foley}, R.~J., \&
 % {Filippenko}, A.~V. 2007, \apj, 661, 1013

\bibitem[{{Matheson} {et~al.}(2008){Matheson}, {Kirshner}, {Challis}, {Jha},
  {Garnavich}, {Berlind}, {Calkins}, {Blondin}, {Balog}, {Bragg}, {Caldwell},
  {Dendy Concannon}, {Falco}, {Graves}, {Huchra}, {Kuraszkiewicz}, {Mader},
  {Mahdavi}, {Phelps}, {Rines}, {Song}, \& {Wilkes}}]{matheson08}
{Matheson}, T., et~al.\ 2008, \aj, 135,  1598

\bibitem[{{Maund} {et~al.}(2005){Maund}, {Smartt}, \& {Schweizer}}]{maund05}
{Maund}, J.~R., {Smartt}, S.~J., \& {Schweizer}, F. 2005, \apj, 630, L33

\bibitem[{{McGaugh}(1991)}]{mcgaugh91}
{McGaugh}, S.~S. 1991, \apj, 380, 140

\bibitem[{{Modjaz} {et~al.}(2008){Modjaz}, {Kewley}, {Kirshner}, {Stanek},
  {Challis}, {Garnavich}, {Greene}, {Kelly}, \& {Prieto}}]{modjaz08_Z}
{Modjaz}, M., et~al.\ 2008,  \aj, 135, 1136

\bibitem[{{Modjaz} {et~al.}(2009){Modjaz}, {Li}, {Butler}, {Chornock},
  {Perley}, {Blondin}, {Bloom}, {Filippenko}, {Kirshner}, {Kocevski},
  {Poznanski}, {Hicken}, {Foley}, {Stringfellow}, {Berlind}, {Barrado y
  Navascues}, {Blake}, {Bouy}, {Brown}, {Challis}, {Chen}, {de Vries},
  {Dufour}, {Falco}, {Friedman}, {Ganeshalingam}, {Garnavich}, {Holden},
  {Illingworth}, {Lee}, {Liebert}, {Marion}, {Olivier}, {Prochaska},
  {Silverman}, {Smith}, {Starr}, {Steele}, {Stockton}, {Williams}, \&
  {Wood-Vasey}}]{modjaz09}
{Modjaz}, M., et~al.\ 2009, \apj, 702, 226

\bibitem[{{Nomoto} {et~al.}(2006){Nomoto}, {Tominaga}, {Umeda}, {Maeda},
  {Ohkubo}, \& {Deng}}]{nomoto06}
{Nomoto}, K., {Tominaga}, N., {Umeda}, H., {Maeda}, K., {Ohkubo}, T., \&
  {Deng}, J. 2006, Nuclear Physics A, 777, 424

\bibitem[{{Oke} {et~al.}(1995){Oke}, {Cohen}, {Carr}, {Cromer}, {Dingizian},
  {Harris}, {Labrecque}, {Lucinio}, {Schaal}, {Epps}, \& {Miller}}]{oke95}
{Oke}, J.~B., et~al.\ 1995, \pasp, 107, 375

\bibitem[{{Pastorello} {et~al.}(2007){Pastorello}, {Smartt}, {Mattila},
  {Eldridge}, {Young}, {Itagaki}, {Yamaoka}, {Navasardyan}, {Valenti}, {Patat},
  {Agnoletto}, {Augusteijn}, {Benetti}, {Cappellaro}, {Boles}, {Bonnet-Bidaud},
  {Botticella}, {Bufano}, {Cao}, {Deng}, {Dennefeld}, {Elias-Rosa},
  {Harutyunyan}, {Keenan}, {Iijima}, {Lorenzi}, {Mazzali}, {Meng}, {Nakano},
  {Nielsen}, {Smoker}, {Stanishev}, {Turatto}, {Xu}, \&
  {Zampieri}}]{pastorello07}
{Pastorello}, A., et~al.\ 2007, \nat, 447, 829

\bibitem[{{P{\'e}rez-Montero} \& {D{\'{\i}}az}(2003)}]{perez03}
{P{\'e}rez-Montero}, E. \& {D{\'{\i}}az}, A.~I. 2003, \mnras, 346, 105

\bibitem[{{Pettini} \& {Pagel}(2004)}]{pettini04}
{Pettini}, M., \& {Pagel}, B.~E.~J. 2004, \mnras, 348, L59

\bibitem[{{Podsiadlowski} {et~al.}(2004){Podsiadlowski}, {Langer},
  {Poelarends}, {Rappaport}, {Heger}, \& {Pfahl}}]{podsiadlowski04}
{Podsiadlowski}, P., {Langer}, N., {Poelarends}, A.~J.~T., {Rappaport}, S.,
  {Heger}, A., \& {Pfahl}, E. 2004, \apj, 612, 1044

\bibitem[{{Prantzos} \& {Boissier}(2003)}]{prantzos03}
{Prantzos}, N., \& {Boissier}, S. 2003, \aap, 406, 259

\bibitem[{{Prieto} {et~al.}(2008){Prieto}, {Stanek}, \& {Beacom}}]{prieto08}
{Prieto}, J.~L., {Stanek}, K.~Z., \& {Beacom}, J.~F. 2008, \apj, 673, 999

\bibitem[{{Rupke} {et al.}(2010)}]{rupke10}
{Rupke}, D., {Kewley}, L. J., \& {Chien}, L.-H.,  2010, \apj, 723, 1255  



\bibitem[{{Sahu} {et~al.}(2009){Sahu}, {Tanaka}, {Anupama}, {Gurugubelli}, \&
  {Nomoto}}]{sahu09}
{Sahu}, D.~K., {Tanaka}, M., {Anupama}, G.~C., {Gurugubelli}, U.~K., \&
  {Nomoto}, K. 2009, \apj, 697, 676

\bibitem[{{Silverman} {et~al.}(2009){Silverman}, {Mazzali}, {Chornock},
  {Filippenko}, {Clocchiatti}, {Phillips}, {Ganeshalingam}, \&
  {Foley}}]{silverman09}
{Silverman}, J.~M., {Mazzali}, P., {Chornock}, R., {Filippenko}, A.~V.,
  {Clocchiatti}, A., {Phillips}, M.~M., {Ganeshalingam}, M., \& {Foley}, R.~J.
  2009, \pasp, 121, 689

\bibitem[{{Smartt}(2009)}]{smartt09_rev}
{Smartt}, S.~J. 2009, \araa, 47, 63

\bibitem[{{Smith} {et~al.}(2011){Smith}, {Li}, {Filippenko}, \&
  {Chornock}}]{smith11_snfrac}
{Smith}, N., {Li}, W., {Filippenko}, A.~V., \& {Chornock}, R. 2011, MNRAS,
  in press (arXiv:1006.3899)

\bibitem[Soderberg et al.(2010)]{soderberg10} Soderberg, A.~M., et 
al.\ 2010, \nat, 463, 513 

\bibitem[{{Starling} {et~al.}(2010){Starling}, {Wiersema}, {Levan}, {Sakamoto},
  {Bersier}, {Goldoni}, {Oates}, {Rowlinson}, {Campana}, {Sollerman}, {Tanvir},
  {Malesani}, {Fynbo}, {Covino}, {D'Avanzo}, {O'Brien}, {Page}, {Osborne},
  {Vergani}, {Barthelmy}, {Burrows}, {Cano}, {Curran}, {De Pasquale}, {D'Elia},
  {Evans}, {Flores}, {Fruchter}, {Garnavich}, {Gehrels}, {Gorosabel}, {Hjorth},
  {Holland}, {van der Horst}, {Jakobsson}, {Kamble}, {Kuin}, {Kaper},
  {Mazzali}, {Nugent}, {Pian}, {Thoene}, \& {Woosley}}]{starling11}
{Starling}, R.~L.~C., et~al.\ 2011, \mnras, in press (arXiv:1004.2919)

\bibitem[{{Stritzinger} {et~al.}(2009){Stritzinger}, {Mazzali}, {Phillips},
  {Immler}, {Soderberg}, {Sollerman}, {Boldt}, {Braithwaite}, {Brown}, {Burns},
  {Contreras}, {Covarrubias}, {Folatelli}, {Freedman}, {Gonz{\'a}lez}, {Hamuy},
  {Krzeminski}, {Madore}, {Milne}, {Morrell}, {Persson}, {Roth}, {Smith}, \&
  {Suntzeff}}]{stritzinger09}
{Stritzinger}, M., et~al.\ 2009, \apj, 696, 713

\bibitem[{{Th{\"o}ne} {et~al.}(2009){Th{\"o}ne}, {Micha{\l}owski}, {Leloudas},
  {Cox}, {Fynbo}, {Sollerman}, {Hjorth}, \& {Vreeswijk}}]{thoene09}
{Th{\"o}ne}, C.~C., {Micha{\l}owski}, M.~J., {Leloudas}, G., {Cox}, N.~L.~J.,
  {Fynbo}, J.~P.~U., {Sollerman}, J., {Hjorth}, J., \& {Vreeswijk}, P.~M. 2009,
  \apj, 698, 1307

\bibitem[{{Tremonti} {et~al.}(2004){Tremonti}, {Heckman}, {Kauffmann},
  {Brinchmann}, {Charlot}, {White}, {Seibert}, {Peng}, {Schlegel}, {Uomoto},
  {Fukugita}, \& {Brinkmann}}]{tremonti04}
{Tremonti}, C.~A., et~al.\ 2004, \apj, 613, 898

\bibitem[{{van Zee} {et~al.}(1998){van Zee}, {Salzer}, {Haynes}, {O'Donoghue},
 \& {Balonek}}]{vanzee98}
{van Zee}, L., {Salzer}, J.~J., {Haynes}, M.~P., {O'Donoghue}, A.~A., \&
 {Balonek}, T.~J. 1998, \aj, 116, 2805

\bibitem[{{Vink} \& {de Koter}(2005)}]{vink05}
{Vink}, J.~S., \& {de Koter}, A. 2005, \aap, 442, 587

\bibitem[{{Woosley} \& {Bloom}(2006)}]{woosley06_rev}
{Woosley}, S.~E., \& {Bloom}, J.~S. 2006, \araa, 44, 507

\bibitem[{{Woosley} {et~al.}(1993){Woosley}, {Langer}, \& {Weaver}}]{woosley93}
{Woosley}, S.~E., {Langer}, N., \& {Weaver}, T.~A. 1993, \apj, 411, 823

\bibitem[{{Young} {et~al.}(2010){Young}, {Smartt}, {Valenti}, {Pastorello},
  {Benetti}, {Benn}, {Bersier}, {Botticella}, {Corradi}, {Harutyunyan},
  {Hrudkova}, {Hunter}, {Mattila}, {de Mooij}, {Navasardyan}, {Snellen},
  {Tanvir}, \& {Zampieri}}]{young10}
{Young}, D.~R., et~al.\ 2010, \aap, 512, A70


\end{thebibliography}
%%%%%%%%%%%%%%%%%%%%%

\clearpage
\clearpage

%%%% TABLE 
%\begin{landscape}

\begin{deluxetable}{lcccccccl}
%\tablewidth{600pt}
\tabletypesize{\scriptsize}
\tablecaption{The Sample of Stripped-Envelope CCSNe}
\tablehead{\colhead{SN Name} & 
\colhead{SN Type} & 
\colhead{SN Host Galaxy} & 
\colhead{Redshift $z$} &
\colhead{Galaxy $M_{B}$} &
\colhead{M91$_{\rm{SNpos}}$} & 
\colhead{KD02-comb$_{\rm{SNpos}}$} & 
\colhead{PP04-O3N2$_{\rm{SNpos}}$} & 
\colhead{SN Disc-type\tablenotemark{a}}\\ 
\colhead{} &
\colhead{} &
\colhead{} &
\colhead{} &
\colhead{[mag]} &
\colhead{log(O/H)+12} &
\colhead{log(O/H)+12} &
\colhead{log(O/H)+12} &
\colhead{} }
\startdata
1990U & Ic & NGC7479 &    0.00794 &  -21.7 & \nodata  &   9.04$^{+ 0.24}_{- 0.09}$ & \nodata  & T \\
1991ar & Ib & IC49 &    0.01521 &  -20.1 &   8.52$^{+ 0.24}_{- 0.11}$ &   8.76$^{+ 0.17}_{- 0.02}$ &   8.56$^{+ 0.05}_{- 0.05}$ & T \\
1996D & Ic & NGC1614 &    0.01582 &  -21.4 &   8.66$^{+ 0.26}_{- 0.06}$ &   8.87$^{+ 0.29}_{- 0.19}$ &   8.59$^{+ 0.03}_{- 0.06}$ & T \\
1996aq & Ib & NGC5584 &    0.00547 &  -19.8 &   8.59$^{+ 0.10}_{- 0.12}$ &   8.92$^{+ 0.04}_{- 0.04}$ &   8.54$^{+ 0.04}_{- 0.04}$ & T \\
1997B & Ic & IC438 &     0.01041 &  -20.7 &   \nodata    & 9.22$^{+ 0.04}_{- 0.04}$ &  $8.94^{+0.13}_{-0.12}$ & T \\
1999cn & Ic & MCG+02-38-043 &    0.02231 &  -19.9 &   8.69$^{+ 0.14}_{- 0.07}$ &   8.76$^{+ 0.14}_{- 0.02}$ &   8.49$^{+ 0.03}_{- 0.04}$ & T \\
1999di & Ib & NGC776 &    0.01641 &  -21.2 &   9.09$^{+ 0.03}_{- 0.03}$ &   9.17$^{+ 0.06}_{- 0.05}$ &   8.94$^{+ 0.07}_{- 0.06}$ & T \\
1999dn & Ib & NGC7714 &    0.00933 &  -20.5 &   8.54$^{+ 0.17}_{- 0.20}$ &   8.39$^{+ 0.13}_{- 0.05}$ &   8.32$^{+ 0.08}_{- 0.09}$ & T \\
2001ig & IIb & NGC7424 &    0.00292 &  -19.6 &   8.27$^{+ 0.07}_{- 0.27}$ &   8.53$^{+ 0.09}_{- 0.11}$ &   8.29$^{+ 0.03}_{- 0.03}$ & T \\
2002bl & Ic-bl & UGC5499 &    0.01591 &  -20.3 &   8.79$^{+ 0.40}_{- 0.39}$ &   9.01$^{+ 0.31}_{- 0.51}$ &   8.65$^{+ 0.29}_{- 0.29}$ & T \\
2004fe & Ic & NGC132 &    0.01788 &  -21.1 &   9.21$^{+ 0.79}_{- 0.91}$ &   8.14$^{+ 0.87}_{- 0.83}$ &   8.65$^{+ 0.00}_{- 0.04}$ & T \\
2004gt & Ic & NGC4038 &    0.00555 &  -21.4 &   8.92$^{+ 0.06}_{- 0.05}$ &   8.99$^{+ 0.14}_{- 0.12}$ &   8.70$^{+ 0.00}_{- 0.01}$ & T \\
2005eo & Ic & UGC04132 &    0.01743 &  -22.2 &   8.49$^{+ 0.51}_{- 0.03}$ &   8.78$^{+ 0.44}_{- 0.24}$ &   8.66$^{+ 0.08}_{- 0.10}$ & T \\
2005mf & Ic & UGC04798 &    0.01891 &  -20.5 &   8.83$^{+ 0.15}_{- 0.36}$ &   8.96$^{+ 0.07}_{- 0.20}$ &   8.66$^{+ 0.09}_{- 0.09}$ & T \\
2006jc & Ib-n & UGC04904 &    0.00548 &  -15.9 & \nodata  &   8.78$^{+ 0.03}_{- 0.03}$ &   8.42$^{+ 0.04}_{- 0.05}$ & T \\
2007cl & Ic & NGC6479 &    0.02218 &  -20.9 & \nodata  &   9.12$^{+ 0.22}_{- 0.05}$ & \nodata  & T \\
2007gr & Ic & NGC1058 &    0.00173 &  -18.6 &   8.37$^{+ 0.58}_{- 0.10}$ &   8.52$^{+ 0.53}_{- 0.32}$ &   8.64$^{+ 0.07}_{- 0.09}$ & T \\
2007rw & IIb & UGC7798 &    0.00857 &  -19.1 &   8.58$^{+ 0.15}_{- 0.03}$ &   8.54$^{+ 0.26}_{- 0.12}$ &   8.38$^{+ 0.03}_{- 0.04}$ & T \\
2007uy & Ib & NGC2770 &    0.00700 &  -20.7 &   8.70$^{+ 0.10}_{- 0.04}$ &   8.83$^{+ 0.11}_{- 0.09}$ &   8.60$^{+ 0.01}_{- 0.02}$ & T \\
2008D & Ib & NGC2770 &    0.00700 &  -20.7 &   8.86$^{+ 0.06}_{- 0.00}$ &   8.81$^{+ 0.07}_{- 0.02}$ &   8.92$^{+ 0.08}_{- 0.06}$ & T \\
2008cx & IIb & NGC309 &    0.01890 &  -22.1 &   7.87$^{+ 1.19}_{- 0.96}$ &   8.50$^{+ 0.75}_{- 0.29}$ &   8.62$^{+ 0.23}_{- 0.29}$ & T \\
2005az & Ic  &  NGC4961  & 0.00875 &   -19.2 &   8.52$^{+0.20}_{-0.20}$ &   8.80$^{+0.08}_{-0.08}$ &   8.62$^{+0.07}_{-0.07}$  & Non-T \\
2005kf & Ic & SDSSJ074726.40+265532.4 &    0.01508 &  -17.0 &   8.70$^{+ 0.08}_{- 0.18}$ &   9.04$^{+ 0.02}_{- 0.07}$ &   8.78$^{+ 0.06}_{- 0.06}$ & Non-T \\
2006fo & Ib & UGC02019 &    0.02074 &  -20.4 &   8.77$^{+ 0.17}_{- 0.13}$ &   8.88$^{+ 0.21}_{- 0.19}$ &   8.75$^{+ 0.01}_{- 0.03}$ & Non-T \\
2006ip & Ic & 2MASXJ23483173-0208524 &    0.03062 &  -19.9 &   8.47$^{+ 0.24}_{- 0.29}$ &   8.85$^{+ 0.07}_{- 0.06}$ &   8.72$^{+ 0.12}_{- 0.09}$ & Non-T \\
2006jo & Ib & SDSSJ012314.96-001948.8 &    0.07678 &  -21.6 &   8.61$^{+ 0.09}_{- 0.09}$ &   8.91$^{+ 0.03}_{- 0.03}$ &   8.48$^{+ 0.03}_{- 0.03}$ & Non-T \\
2006ld & Ib & UGC348 &    0.01394 &  -18.5 &   8.63$^{+ 0.26}_{- 0.03}$ &   8.61$^{+ 0.18}_{- 0.30}$ &   8.23$^{+ 0.14}_{- 0.14}$ & Non-T \\
2006lt & Ib & NSFJ021659.89+304157.4 &    0.01602 & \nodata  &   8.59$^{+ 0.13}_{- 0.11}$ &   8.73$^{+ 0.08}_{- 0.07}$ &   8.57$^{+ 0.06}_{- 0.05}$ & Non-T \\
2007eb & Ic-bl & NSFJ224248.98+240247.2 &    0.04262 & \nodata  &   8.32$^{+ 0.28}_{- 0.33}$ &   8.43$^{+ 0.31}_{- 0.08}$ &   8.26$^{+ 0.07}_{- 0.07}$ & Non-T \\
2007eq & Ib & NSFJ234805.93+281420.2 &    0.02964 & \nodata  &   8.50$^{+ 0.23}_{- 0.18}$ &   8.51$^{+ 0.14}_{- 0.18}$ &   8.34$^{+ 0.09}_{- 0.08}$ & Non-T \\
2007fj   &  Ic & APMUKS B221135.36-2826& 0.05680 & \nodata & \nodata   &   8.85$^{+0.05}_{-0.05}$ &  8.50$^{+0.04}_{-0.04}$  & Non-T \\  %
2007gx & Ic-bl & NSFJ171851.49+224716.6 &    0.07894 & \nodata  & \nodata  &   9.14$^{+ 0.02}_{- 0.02}$ & \nodata  & Non-T \\
2007jy & Ib & SDSSJ205121.43+002357.8 &    0.18295 &  -19.6 &   8.48$^{+ 0.05}_{- 0.05}$ &   8.92$^{+ 0.01}_{- 0.01}$ &   8.41$^{+ 0.02}_{- 0.02}$ & Non-T \\
2007qw & Ic-bl & SDSSJ223529.00+002856.1 &    0.15064 &  -19.4 &   8.46$^{+ 0.06}_{- 0.04}$ &   8.50$^{+ 0.10}_{- 0.06}$ &   8.19$^{+ 0.00}_{- 0.01}$ & Non-T \\
2008cw & IIb & SDSSJ163238.15+412730.7 &    0.03193 &  -18.3 & \nodata  &   8.46$^{+ 0.01}_{- 0.01}$ & \nodata  & Non-T \\
\\
\hline
\\
1997ef\tablenotemark{b} & Ic-bl & UGC4107 &    0.01169 &  -20.2 &   8.84$^{+ 0.02}_{- 0.03}$ &   8.93$^{+ 0.04}_{- 0.04}$ &   8.69$^{+ 0.02}_{- 0.03}$ & T \\
1998ey\tablenotemark{b} & Ic-bl & NGC7080 &    0.01610 &  -21.8 & \nodata  &   9.08$^{+ 0.04}_{- 0.04}$ & \nodata  & T \\
2002ap\tablenotemark{b} & Ic-bl & NGC628 &    0.00220 &  -20.6 &   8.56$^{+ 0.07}_{- 0.07}$ &   8.62$^{+ 0.06}_{- 0.06}$ &   8.38$^{+ 0.06}_{- 0.06}$ & T \\
2003jd\tablenotemark{b} & Ic-bl & MCG-01-59-21-20.3 &    0.01880 &  -20.3 &   8.62$^{+ 0.02}_{- 0.02}$ &   8.59$^{+ 0.05}_{- 0.05}$ &   8.39$^{+ 0.02}_{- 0.02}$ & T \\
2007ru\tablenotemark{c} & Ic-bl & UGC12381 &    0.02218 &  -20.3 & \nodata  &   8.98$^{+ 0.04}_{- 0.04}$ & \nodata  & T \\
2007Y\tablenotemark{d} & Ib & NGC1187 &    0.00464 &  -20.2 &   8.14$^{+ 0.13}_{- 0.07}$ &   8.51$^{+ 0.23}_{- 0.15}$ &   8.25$^{+ 0.01}_{- 0.02}$ & T \\
2005kr\tablenotemark{b} & Ic-bl & SDSSJ030829.66+005320.1 &    0.13450 &  -17.4 &   8.64$^{+ 0.01}_{- 0.01}$ &   8.63$^{+ 0.03}_{- 0.03}$ &   8.24$^{+ 0.01}_{- 0.01}$ & Non-T \\
2005ks\tablenotemark{b} & Ic-bl & SDSSJ213756.52-000157.7 &    0.09870 &  -19.2 &   8.71$^{+ 0.03}_{- 0.04}$ &   8.87$^{+ 0.02}_{- 0.03}$ &   8.63$^{+ 0.01}_{- 0.01}$ & Non-T \\
2005nb\tablenotemark{b} & Ic-bl & UGC7230 &    0.02377 &  -21.3 &   8.56$^{+ 0.03}_{- 0.03}$ &   8.65$^{+ 0.08}_{- 0.09}$ &   8.49$^{+ 0.03}_{- 0.03}$ & Non-T \\
2006nx\tablenotemark{b} & Ic-bl & SDSSJ033330.43-004038.0 &    0.13700 &  -18.9 &   8.48$^{+ 0.09}_{- 0.10}$ &   8.53$^{+ 0.10}_{- 0.14}$ &   8.24$^{+ 0.04}_{- 0.03}$ & Non-T \\
2006qk\tablenotemark{b} & Ic-bl & SDSSJ222532.38+000914.9 &    0.05840 &  -17.9 &   8.61$^{+ 0.07}_{- 0.08}$ &   8.82$^{+ 0.04}_{- 0.05}$ &   8.75$^{+ 0.02}_{- 0.03}$ & Non-T \\
2007I\tablenotemark{b} & Ic-bl & SDSSJ115913.13-013616.1 &    0.02160 &  -16.9 & \nodata  &   8.71$^{+ 0.03}_{- 0.04}$ &   8.39$^{+ 0.40}_{- 0.17}$ & Non-T \\

\enddata
\tablecomments{SNe above the horizontal line: Local host-galaxy
  spectra are presented and analyzed here for the first time. SNe
  below the horizontal line: we measured line fluxes and computed
  oxygen abundances from previously published spectra.}
\tablenotetext{a}{SN Discovery Type: T = SN host galaxy was targeted;
  Non-T = SN host galaxy was not targeted.}
\tablenotetext{b}{Sample and data from \citet{modjaz08_Z}, which uses
  the same technique as this work.}
\tablenotetext{c}{Remeasured from spectrum in  \citet{sahu09}.}
\tablenotetext{d}{Measured from spectrum in \citet{stritzinger09}.}
\label{sample_table}
\end{deluxetable}

\end{document}